\documentclass[preprint,aps]{revtex4}
\usepackage{graphicx} % For eps figures
\usepackage{color}
\usepackage{amsmath}
\begin{document}

\title{New uncertainty relations for
tomographic entropy: Application to squeezed states and solitons}

\author{{\bf Sergio De Nicola}\footnote{Email: s.denicola@cib.na.cnr.it}}
\affiliation{\small Istituto di Cibernetica ``Eduardo Caianiello''
del CNR Comprensorio ``A. Olivetti'' Fabbr. 70, Via Campi Flegrei,
34, I-80078 Pozzuoli (NA), Italy}
\author{{\bf Renato Fedele}\footnote{Email: renato.fedele@na.infn.it}}
\affiliation{\small Dipartimento di Scienze Fisiche,
Universit\`{a} Federico II and INFN Sezione di Napoli, Complesso
Universitario di M.S. Angelo, via Cintia, I-80126 Napoli, Italy}
\author{{\bf Margarita A. Man'ko\footnote{Email: mmanko@sci.lebedev.ru}}
and {\bf Vladimir I. Man'ko}\footnote{Emails: manko@na.infn.it~~
manko@sci.lebedev.ru}} \affiliation{\small P. N. Lebedev Physical
Institute, Leninskii Prospect 53, Moscow 119991 Russia}

\begin{abstract}
Using the tomographic probability distribution (symplectic tomogram)
describing the quantum state (instead of the wave function or
density matrix) and properties of recently introduced tomographic
entropy associated with the probability distribution, the new
uncertainty relation for the tomographic entropy is obtained.
Examples of the entropic uncertainty relation for squeezed states
and  solitons of the Bose--Einstein condensate are considered.

\medskip
\noindent
PACS number(s): 42.50.-p, 42.50.Dv, 03.67.-a

\begin{center}
{ To be published in European Physical Journal B}
\end{center}

\end{abstract}

\maketitle

\section{Introduction}\label{sect1}

Quantum mechanics is known to differ from classical mechanics due
to the existence of the position--momentum uncertainty relation by
Heisenberg~\cite{Heis,Kenn}. The uncertainty relation containing
the correlation of the position and momentum was found by
Robertson~\cite{Rob29} and
Schr\"odinger~\cite{Schr30,Kur80,SudPLA}. There exists the
uncertainty relation of the position and momentum for mixed
quantum states~\cite{183}. New kinds of the uncertainty relations
were obtained by Trifonov~\cite{Tri1-3}. Extensions of the
uncertainty relations of \cite{Rob35} for mixed states were found
by Karelin~\cite{Kar}. Review of the uncertainty relations in
quantum mechanics is given in \cite{Dodonov,Ozawa}.

There exist specific uncertainty relations called ``entropic
uncertainty relations'' based on the notion of Shannon entropy and
information~\cite{Shannon}. These relations, which read as
inequalities for entropy associated with the position-and-momentum
probability distributions, were discussed, for example, in
\cite{Bal-Bir,Hirsh}.

Recently a new formulation of quantum mechanics where the quantum
states are described by tomographic-probability distributions
(instead of the wave function or density matrices) was
suggested~\cite{ManciniPLA-FP}. For a system with continuous
degrees of freedom, such probability is the symplectic tomogram of
the quantum state~\cite{Dariano-Mancini96}. The corresponding
symplectic tomographic entropy was introduced for quantum states
in \cite{Olga97} and in signal analysis in \cite{Rita}. In
\cite{EurPJ-Renato} the symplectic entropy was discussed for the
BEC solitons, in view of the tomogram of the solution to
Gross--Pitaevskii equation. A general approach to quantum
information including the application of different kinds of
tomographic entropies was developed in \cite{Rui-ArXiv}.

The aim of this study is to establish a new kind of entropic
uncertainty relations formulated as inequality for the entropy
associated with the symplectic tomogram of the quantum state of a
system with continuous degrees of freedom.

The paper is organized as follows.

In section \ref{sect2}, a review of known entropic uncertainty
relations for systems with continuous variables is presented
while, in section \ref{sect3}, the symplectic-tomography approach
is discussed. Entropic inequalities for symplectic entropy are
studied in section \ref{sect4} and examples of new inequalities
for the gaussian packets (squeezed states) and soliton solution of
Gross--Pitaevskii equation  (Bose--Einstein condensates) are given
in section \ref{sect5}. Finally, conclusions are summarized in
section \ref{sect6}.

\section{Entropy and Entropic Uncertainty Relations}\label{sect2}

In the context of information theory, entropy is related to an
arbitrary probability-distribution function~\cite{Shannon}. For
example, given the probability distribution $P(n)$, where $n$ is a
discrete random variable, i.e.,
\begin{equation}\label{A1}
P(n)\geq 0,
\end{equation}
together with the normalization condition
\begin{equation}\label{A2}
\sum_nP(n)=1,
\end{equation}
one has, by definition, the entropy
\begin{equation}\label{A3}
S=-\sum_nP(n)\ln P(n)=-\langle\ln P(n)\rangle.
\end{equation}

In quantum mechanics, the discrete probability distributions are
standard ingredients in the description of spin (see, e.g.,
tomographic probability of spin states in \cite{DodPLA,OlgaJETP}
and related entropy for spin tomograms in \cite{OlgaJRLR}).

For continuous variables, the wave function $\psi(x)$ provides the
probability-distribution density
\begin{equation}\label{A4}
P(x)=|\psi(x)|^2.
\end{equation}
The corresponding entropy reads (see, e.g., \cite{183})
\begin{equation}\label{A5}
S_x=-\int|\psi(x)|^2\ln|\psi(x)|^2\,dx.
\end{equation}

In the momentum representation, one has the wave function
\begin{equation}\label{A6}
\widetilde\psi(p)=\frac{1}{\sqrt{2\pi}}\int\psi(x)e^{-ipx}\,dx
\qquad (\hbar=1).
\end{equation}
The corresponding entropy related to the momentum-probability
density $|\widetilde\psi(p)|^2$ reads
\begin{equation}\label{A7}
S_p=-\int|\widetilde\psi(p)|^2\ln|\widetilde\psi(p)|^2\,dp.
\end{equation}

It is worthy noting that one can construct entropies $S_x$ and
$S_p$ not only in quantum mechanics. If the function $\psi(x)$ is
replaced by a signal function $f(t)$ depending on time $t$, the
function $\widetilde\psi(p)$ is replaced by the function
$\widetilde f(\omega)$ describing the signal spectrum.

In this case, the entropy of the signal
\begin{equation}\label{A8}
S_t=-\int|f(t)|^2\ln|f(t)|^2\,dt
\end{equation}
and its spectrum
\begin{equation}\label{A9}
S_\omega=-\int|\widetilde f(\omega)|^2\ln|\widetilde
f(\omega)|^2\,d\omega
\end{equation}
provide some information characteristics of the signal.

From mathematical point of view, there exists the correlation of
entropies $S_x$ and $S_p$ $(S_t$ and $S_\omega)$, since the
function $\psi(x)$ $[f(t)]$ determines the Fourier component
$\widetilde\psi(p)$ $[\widetilde f(\omega)]$. This means that the
entropies $S_x$ and $S_p$ have to obey some constrains. These
constrains are entropic uncertainty relations (some inequalities).

For the one-mode system, the inequality reads (see \cite{183},
p.~28)
\begin{equation}\label{A10}
S_x+S_p\geq\ln(\pi e),
\end{equation}
or
\begin{equation}\label{A11}
S_t+S_\omega\geq\ln(\pi e).
\end{equation}

For the Gaussian wave functions (Gaussian signals) describing the
states without correlations of the position and momentum, e.g.,
the ground state of the harmonic oscillator
\begin{equation}\label{A12}
\psi(x)=\pi^{-1/4}e^{-x^2/2},\qquad\widetilde\psi(p)
=\pi^{-1/4}e^{-p^2/2},
\end{equation}
one has
\begin{equation}\label{A13}
S_x^{(0)}=S_p^{(o)}=\frac{1}{2}\ln(\pi e).
\end{equation}
Consequently,
\begin{equation}\label{A14}
S_x^{(0)}+S_p^{(0)}=\ln(\pi e).
\end{equation}
The equality takes place for squeezed states with the wave function
\begin{equation}\label{A12ss}
\psi(x)=\left(2\pi\sigma_{x^2}\right)^{-1/4}e^{-x^2/4\sigma_{x^2}}.
\end{equation}
Thus, one has
\begin{equation}\label{A15}
S_x=\frac{1}{2}\ln(2\pi e\sigma_{x^2}),\qquad
S_p=\frac{1}{2}\ln(2\pi e\sigma_{p^2}),
\end{equation}
where $\sigma_{x^2}$ and $\sigma_{p^2}$ read
\begin{eqnarray}\label{A16}
\sigma_{x^2}&=&\int x^2|\psi(x)|^2\,dx-\left(\int
x|\psi(x)|^2\,dx\right)^2,\nonumber\\ &&\\ \sigma_{p^2}&=&\int
p^2|\widetilde\psi(p)|^2\,dp-\left(\int
p|\widetilde\psi(p)|^2\,dp\right)^2,\nonumber
\end{eqnarray}
and
\begin{equation}\label{A17}
\sigma_{x^2}\sigma_{p^2}=\frac{1}{4}\,.
\end{equation}

For squeezed and correlated~\cite{Kur80} states, the wave
functions have the Gaussian form, i.e., $$\psi(x)={\cal
N}\exp(-ax^2+bx),\qquad a=a_1+ia_2,$$ and
\begin{equation}\label{A18}
\sigma_{x^2}\sigma_{p^2}=\frac{1}{4}\,\frac{1}{1-R^2}\,.
\end{equation}
Here $R$ is the correlation coefficient of the position and
momentum, i.e.,
\begin{equation}\label{A19}
R=\frac{\frac{1}{2}\langle\hat q\hat p+\hat p\hat
q\rangle-\langle\hat q\rangle\langle\hat p\rangle}{\sqrt{
\sigma_{x^2}\sigma_{p^2}}}\,,\qquad |R|<1,
\end{equation}
and for squeezed but not correlated states $R=0$.

The sum of entropies for the squeezed and correlated states reads
\begin{equation}\label{A20}
S_x+S_p=\ln(\pi e)+\ln\frac{1}{\sqrt{1-R^2}}\geq\ln(\pi e).
\end{equation}
For squeezed but not correlated states, the entropy $S_x$ differs
from $S_p$.

For multimode systems (multicomponent signals), the entropy
uncertainty relation reads
\begin{equation}\label{A21}
S_{\vec x}+S_{\vec p}\geq N\ln(\pi e),
\end{equation}
where $N$ is the number of degrees of freedom of the system and
\begin{eqnarray}\label{A22}
S_{\vec x}&=&-\int|\psi(\vec x)|^2\ln|\psi(\vec x)|^2\,d\vec x,
\nonumber\\ &&\\ S_{\vec p}&=&-\int|\widetilde\psi(\vec
p)|^2\ln|\widetilde\psi(\vec p)|^2\,d\vec p.
\nonumber\end{eqnarray} The functions $\psi(\vec x)$ and
$\widetilde\psi(\vec p)$ are connected by the Fourier transform
\begin{equation}\label{A23}
\widetilde\psi(\vec p)=(2\pi)^{-N/2}\int\psi(\vec x)e^{-i\vec
p\vec x}\,d\vec x.
\end{equation}

For the Gaussian wave function corresponding to factorized
squeezed state of several modes,
\begin{equation}\label{A24}
S_{\vec x}+S_{\vec p}= N\ln(\pi e).
\end{equation}

\section{Symplectic Tomography}\label{sect3}

There exists an invertable map of the density operator (matrix)
$\hat\rho$ onto the symplectic tomogram~\cite{Mancini95,OlgaJPA}
which is the probability-density of random quadrature $X$
\begin{equation}\label{T1}
w(X,\mu,\nu)=\mbox{Tr}\,\hat\rho\,\,\delta(X-\mu\hat q-\nu\hat p).
\end{equation}
Here $\hat\rho$ is the density operator. The parameters $\mu$ and
$\nu$ are real parameters and operators $\hat q$ and $\hat p$ are
quadrature operators.

The map (\ref{T1}) has the inverse~\cite{Dariano-Mancini96}
\begin{equation}\label{T3}
\hat\rho=\frac{1}{2\pi}\int w(X,\mu,\nu)\exp\left[i\left(X-\mu\hat
q-\nu\hat p\right)\right]dX\,d\mu\,d\nu.
\end{equation}

For a pure state $\hat\rho_\psi=\mid\psi\rangle\langle\psi\mid$,
the transform (\ref{T1}) yields~\cite{Mendes}
\begin{equation}\label{T4}
w(X,\mu,\nu)=\frac{1}{2\pi|\nu|}\left|\int\psi(y)\exp\left(
\frac{i\mu}{2\nu}y^2-\frac{iX}{\nu}y\right)dy\right|^2.
\end{equation}

The function $w(X,\mu,\nu)$ (the state tomogram) is the
probability density of the position $X$, i.e.,
$$w(X,\mu,\nu)\geq 0$$ and
\begin{equation}\label{T5}
\int w(X,\mu,\nu)\,dX=1.
\end{equation}

Transformation (\ref{T1}) can be expressed in terms of the real
Wigner function~\cite{Wigner32}
\begin{equation}\label{T6}
W(q,p)=\int\rho\left(q+\frac{u}{2}\,,q-\frac{u}{2}\right)e^{-ipu}\,du,
\end{equation}
where $\rho(x,x')$ is the density matrix in the position
representation, and for $\mbox{Tr}\,\hat\rho=1$ one has
\begin{equation}\label{T7}
\int W(q,p)\,\frac{dq\,dp}{2\pi}=1.
\end{equation}
The density matrix reads
\begin{equation}\label{T8A}
\rho(x,x')=\frac{1}{2\pi}\int W
\left(\frac{x+x'}{2}\,,p\right)e^{ip(x-x')}\,dp.
\end{equation}
In terms of the tomogram, one has
\begin{equation}\label{T8}
\rho(x,x')=\frac{1}{2\pi}\int
w(Y,\mu,x-x')e^{-i[Y-\mu(x+x')/2]}\,dY\,d\mu.
\end{equation}
The tomographic-probability density has the homogeneity property
following from its definition (\ref{T1}) and the relation for the
Dirac delta-function
\begin{equation}\label{T9}
\delta(\lambda y)=\frac{1}{|\lambda|}\delta(y).
\end{equation}
The homogeneity property reads
\begin{equation}\label{T10}
w(\lambda
X,\lambda\mu,\lambda\nu)=\frac{1}{|\lambda|}w(X,\mu,\nu).
\end{equation}
Also for the pure state, one has
\begin{equation}\label{T11}
w(X,1,0)=|\psi(X)|^2
\end{equation}
and
\begin{equation}\label{T12}
w(X,0,1)=|\widetilde\psi(X)|^2,
\end{equation}
where $\psi(X)$ is the wave function in the position
representation and $\widetilde\psi(X)$ is the wave function in the
momentum representation.

These properties are connected with the expression of symplectic
tomogram in terms of the Wigner function~\cite{Mancini95}
\begin{equation}\label{T13}
w(X,\mu,\nu)=\int W(q,p)\,\delta(X-\mu q-\nu
p)\,\frac{dq\,dp}{2\pi}\,.
\end{equation}
The inverse transform reads
\begin{equation}\label{T14}
W(q,p)=\frac{1}{2\pi}\int w(X,\mu,\nu)\exp\left[i\left(X-\mu q-\nu
p\right)\right]\,dX\,d\mu\,d\nu.
\end{equation}

Since for the pure state,
\begin{equation}\label{T15}
\int W(q,p)\frac{dp}{2\pi}=|\psi(q)|^2
\end{equation}
and
\begin{equation}\label{T16}
\int W(q,p)\frac{dq}{2\pi}=|\widetilde\psi(p)|^2,
\end{equation}
relations (\ref{T11}) and (\ref{T12}) are easily obtained.

Tomogram (\ref{T4}) can be rewritten in the form
\begin{equation}\label{T17}
w(X,\mu,\nu)=\frac{1}{2\pi|\nu|}\left|\int\psi(y)\exp\left[
\frac{i}{2}\left(\frac{\mu}{\nu}y^2-\frac{2X}{\nu}y+
\frac{\mu}{\nu}X^2\right)\right]dy\right|^2.
\end{equation}
For $\mu=\cos t$ and $\nu=\sin t$, one has the optical
tomogram~\cite{BerBer,VogRis}
\begin{equation}\label{T18}
w(X,t)=\left|\int\psi(y)\exp\left[\frac{i}{2}\left(\mbox{cot}\,t\,\,
(y^2+X^2)-\frac{2X}{\sin t}\,y\right)\right]\frac{dy}{\sqrt{2\pi
i\sin t}}\right|^2.
\end{equation}
On the other hand, this tomogram formally equals to
\begin{equation}\label{T21}
w(X,t)=|\psi(X,t)|^2,
\end{equation}
where the wave function reads
\begin{equation}\label{T22}
\psi(X,t)=\frac{1}{\sqrt{2\pi i\sin t}}\int
\exp\left[\frac{i}{2}\left(\mbox{cot}\,t\,\,
(y^2+X^2)-\frac{2X}{\sin t}\,y\right)\right]\psi(y)\,{dy},
\end{equation}
being the fractional Fourier transform of the wave function
$\psi(y)$. This wave function corresponds to the wave function of
a harmonic oscillator with $\hbar=m=\omega=1$ taken at the time
moment $t$ provided the wave function at the initial time moment
$t=0$ equals to $\psi(y)$.

For mixed state with density operator $\hat\rho$ written in the
form of the spectral decomposition,
\begin{equation}\label{T23}
\hat\rho=\sum_k\lambda_k\mid\psi_k\rangle\langle\psi_k\mid,
\end{equation}
where $\lambda_k$ are nonnegative eigenvalues and
$\mid\psi_k\rangle$ are the eigenvectors of the density operator,
the optical tomogram reads
\begin{equation}\label{T24}
w(X,\mu=\cos t,\nu=\sin t)=\sum_k\frac{\lambda_k}{2\pi |\sin
t|}\left|\psi_k(y)\exp\left[\frac{i}{2}\left(\mbox{cot}\,t\,\,
(y^2+X^2)-\frac{2X}{\sin t}\right)\right]{dy}\right|^2.
\end{equation}
In Eqs. (\ref{T22}) and (\ref{T24}), we use the identity of the
kernel of fractional Fourier transform to the Green function of
the Schr\"odinger evolution equation for the harmonic oscillator
\cite{RitaJRLR20-99-226}.

Tomogram of a mixed state takes the form of convex sum of
tomograms of pure states $\mid\psi_k\rangle$, i.e.,
\begin{equation}\label{T25}
w(X,\mu,\nu)=\sum_k\lambda_kw_k(X,\mu,\nu),
\end{equation}
where $w_k(X,\mu,\nu)$ are given by Eq. (\ref{T17}) with
$$\psi(y)\rightarrow\psi_k(y)=\langle y\mid\psi_k\rangle.$$

In view of Eqs. (\ref{T11}) and (\ref{T12}), one has for mixed
state
\begin{equation}\label{T26}
w(X,1,0)=\sum_k\lambda_k|\psi_k(X)|^2
\end{equation}
and
\begin{equation}\label{T27}
w(X,0,1)=\sum_k\lambda_k|\widetilde\psi_k(X)|^2,
\end{equation}
where $\psi_k(X)$ is the complex wave function in the position
representation of the eigenstate $\mid\psi_k\rangle$ and
$\widetilde\psi_k(X)$ is the complex wave function in the momentum
representation of this state.

Thus, we pointed out that symplectic tomogram of a quantum state
can be interpreted as modulus squared of the harmonic-oscillator's
wave function for pure state or as convex sum of modulus squared
of such functions for mixed state.

Another possibility for analogous interpretation follows in view
of considering the tomogram $w(X,\mu,\nu)$ within the framework of
Fresnel-tomography approach \cite{DeNicola}.

In fact, formula (\ref{T4}) with $\mu=1$ can be rewritten in the
form
\begin{equation}\label{T28}
w_{\rm F}(X,\nu)\equiv w(X,\mu=1,\nu)=\left|\int\frac{1}{\sqrt{2\pi
i\nu}}\exp\left[ \frac{i(X-y)^2}{2\nu}\right]\psi(y)\,dy\right|^2.
\end{equation}

The Fresnel tomogram $w_{\rm F}\left(X,\nu\right)$ is related to the
optical tomogram by
\begin{equation}\label{T30}
w_{\rm F}\left(\frac{X}{\mu}\,,
\frac{\nu}{\mu}\right)=|\mu|w(X,\mu,\nu).
\end{equation}
On the other hand, $w_{\rm F}\left({X}/{\mu},{\nu}/{\mu}\right)$ can
be considered as the wave function of free particle at the time
moment $t=\nu$ if the initial value of the wave function at the time
moment $t=0$ is equal to $\psi(y)$.

Thus, the Fresnel tomogram for pure state can be interpreted as
modulus squared of the wave function of free particle.

The Fresnel tomogram is the probability distribution satisfying the
normalization condition
\begin{equation}\label{T31}
\int w_{\rm F}(X,\nu)\,dX=1.
\end{equation}
For mixed state (\ref{T23}), the Fresnel tomogram reads
\begin{equation}\label{T32}
w_{\rm F}(X,\nu)=\sum_k\lambda_k\left|\frac{1}{\sqrt{2\pi
i\nu}}\int\exp\left[\frac{i(X-y)^2}{2\nu}\right]\psi_k(y)\,dy\right|^2
\end{equation}
and
\begin{equation}\label{T32a}
w_{\rm F}(X,0)=\sum_k\lambda_k|\psi_k(X)|^2.
\end{equation}

\section{Tomographic Entropies}\label{sect4}

Since the symplectic tomogram has the standard probability
distribution features, one can introduce entropy associated with
the tomogram of quantum state \cite{Olga97} or of analytic signal
\cite{Rita}. Thus one has entropy as the function of two real
variables
\begin{equation}\label{T33}
S(\mu,\nu)=-\int w(X,\mu,\nu)\,\ln w(X,\mu,\nu)\,dX.
\end{equation}
In view of the homogeneity and normalization conditions for
tomogram (\ref{T10}), (\ref{T5}) one has the additivity property
\begin{equation}\label{T34}
S(\lambda\mu,\lambda\nu)=S(\mu,\nu)+\ln |\lambda|.
\end{equation}
For pure state $\mid\psi\rangle$, one obtains the entropies $S_x$
and $S_p$, namely,
\begin{equation}\label{T35}
S(1,0)=S_x
\end{equation}
and
\begin{equation}\label{T36}
S(0,1)=S_p.
\end{equation}
In view of inequality (\ref{A10}), one has the inequality for
tomographic entropies
\begin{equation}\label{T37}
S(1,0)+S(0,1)\geq \ln (\pi e).
\end{equation}
For multimode system, the symplectic entropy reads
\begin{equation}\label{T38}
S(\vec\mu,\vec\nu)=-\int w(\vec X,\vec\mu,\vec\nu)\,\ln w(\vec
X,\vec\mu,\vec\nu)\,d\vec X.
\end{equation}
Since the symplectic entropy is related to entropies $S_{\vec x}$
and $S_{\vec p}$ of the multimode-system state, one can use
inequality (\ref{A21}) to obtain the entropic uncertainty relation
in the form of inequality for symplectic entropies
\begin{equation}\label{T39}
S(\vec 1,\vec 0)+S(\vec 0,\vec 1)\geq N\ln(\pi e),
\end{equation}
where $\vec\mu=\vec 1$ means $\vec\mu=(1,1,\ldots,1)$ and
$\vec\nu=\vec 1$ means $\vec\nu=(1,1,\ldots,1)$.

The Fresnel tomogram provides the Fresnel entropy of the quantum
state
\begin{equation}\label{T40}
S_{\rm F}(\nu)=-\int w_{\rm F}(X,\nu)\,\ln w_{\rm F}(X,\nu)\,dX.
\end{equation}
It can be readily seen that the Fresnel entropy $S_{\rm F}(\nu)$ can
be easily obtained from the symplectic entropy (\ref{T33}) choosing
$\mu=1$, i.e.,
\begin{equation}\label{T41}
S(1,\nu)=S_{\rm F}(\nu).
\end{equation}
This also means that
\begin{equation}\label{T42}
S_{\rm F}(0)=S_x.
\end{equation}

For the optical tomogram (\ref{T18}), entropy is defined by the
formula
\begin{equation}\label{T43}
S(t)=-\int w(X,t)\,\ln w(X,t)\,dX.
\end{equation}
For the pure state, one has
\begin{equation}\label{T44}
S(0)=S_x
\end{equation}
and
\begin{equation}\label{T45}
S(\pi/2)=S_p.
\end{equation}
In view of the expression of tomogram in terms of wave function
(\ref{T21}) and (\ref{T22}), one has the entropic uncertainty
relation in the form
\begin{equation}\label{T46}
S(t)+S(t+\pi/2)\geq\ln \pi e.
\end{equation}
Since symplectic and optical tomograms are connected as follows:
\begin{equation}\label{T47}
w(X,\mu=\cos t,\nu=\sin t)=w(X,t),
\end{equation}
the corresponding entropies are also connected
\begin{equation}\label{T48}
S(t)=S(\mu=\cos t,\nu=\sin t).
\end{equation}
For given symplectic entropy of any pure state $S(\mu,\nu)$,
inequality (\ref{T46}) reads
\begin{equation}\label{T49}
S(\cos t,\sin t)+S(-\sin t,\cos t)\geq\ln \pi e.
\end{equation}
The optical tomogram $w(x,t)$ and symplectic tomogram
$w(X,\mu,\nu)$ connected by (\ref{T47}) can be related by another
formula
\begin{equation}\label{T50}
w(X,\mu,\nu)=\frac{1}{\sqrt{\mu^2+\nu^2}}\,w\left(
\frac{X}{\sqrt{\mu^2+\nu^2}}\,,t\right).
\end{equation}
This means that for given optical tomogram $w(x,t)$ one can
reconstruct symplectic tomogram $w(X,\mu,\nu)$. Inserting Eq.
(\ref{T50}) into basic equation defining the entropy (\ref{T33})
yields the equality
\begin{equation}\label{T51}
S(t)=S\Big(\sqrt{\mu^2+\nu^2}\cos t,\sqrt{\mu^2+\nu^2}\sin t\Big)
-\frac{1}{2}\,\ln(\mu^2+\nu^2).
\end{equation}

For symplectic entropies (\ref{T49}), the entropic uncertainty
relation yields
\begin{eqnarray}\label{T52}
&&S\Big(\sqrt{\mu^2+\nu^2}\cos t,\sqrt{\mu^2+\nu^2}\sin t\Big)+
S\Big(-\sqrt{\mu^2+\nu^2}\sin t,\sqrt{\mu^2+\nu^2}\cos
t\Big)\nonumber\\&&-\ln(\mu^2+\nu^2)\geq\ln \pi e.
\end{eqnarray}
The extension of this inequality for multimode system reads
\begin{eqnarray}
&&S\Big(\sqrt{\mu_1^2+\nu_1^2}\cos t_1,\sqrt{\mu_2^2+\nu_2^2}\cos
t_2,\ldots,\sqrt{\mu_N^2+\nu_N^2}\cos t_N,\nonumber\\
&&\sqrt{\mu_1^2+\nu_1^2}\sin t_1,\sqrt{\mu_2^2+\nu_2^2}\sin
t_2,\ldots,\sqrt{\mu_N^2+\nu_N^2}\sin t_N\Big)\nonumber\\ &&
+S\Big(-\sqrt{\mu_1^2+\nu_1^2}\sin t_1,-\sqrt{\mu_2^2+\nu_2^2}\sin
t_2,\ldots,-\sqrt{\mu_N^2+\nu_N^2}\sin t_N,\nonumber\\&&
\sqrt{\mu_1^2+\nu_1^2}\cos t_1,\sqrt{\mu_2^2+\nu_2^2}\cos
t_2,\ldots,\sqrt{\mu_N^2+\nu_N^2}\cos t_N\Big)\nonumber\\&&
-\sum_{k=1}^N\ln\left(\mu_k^2+\nu_k^2\right)\geq N\ln(\pi e),
\label{T53}\end{eqnarray} where entropy $S(\vec\mu,\vec\nu)$ is
given by (\ref{T38}).

Tomogram of the ground state of multimode isotropic harmonic
oscillator with unit masses and frequencies has the form
\begin{equation}\label{T54}
w_0\left(\vec X,\vec\mu,\vec\nu\right)=\prod_{k=1}^N\frac{1}
{\sqrt{\pi\left(\mu_k^2+\nu_k^2\right)}}\,\exp\left(-\frac{X_k^2}
{\mu_k^2+\nu_k^2}\right).
\end{equation}
Entropy associated with this tomogram reads
\begin{equation}\label{T55}
S_0\left(\vec\mu,\vec\nu\right)=\frac{N}{2}\,\ln\pi+\frac{N}{2}+
\frac{N}{2}\sum_{k=1}^N\ln\left(\mu_k^2+\nu_k^2\right).
\end{equation}
This entropy does not depend on the parameter $t_k$.

One can check that, if $\mu_k\rightarrow\sqrt{\mu_k^2+\nu_k^2}\cos
t_k$ and $\nu_k\rightarrow\sqrt{\mu_k^2+\nu_k^2}\sin t_k$ in
formula (\ref{T55}), relation (\ref{T53}) yields for $S_0$ the
equality
\begin{eqnarray}
&& S_0\Big(\sqrt{\mu_1^2+\nu_1^2}\cos
t_1,\sqrt{\mu_2^2+\nu_2^2}\cos
t_2,\ldots,\sqrt{\mu_N^2+\nu_N^2}\cos t_N,\nonumber\\&&
\sqrt{\mu_1^2+\nu_1^2}\sin t_1,\sqrt{\mu_2^2+\nu_2^2}\sin
t_2,\ldots,\sqrt{\mu_N^2+\nu_N^2}\sin t_N\Big)\nonumber\\&&
+S\Big(-\sqrt{\mu_1^2+\nu_1^2}\sin t_1,-\sqrt{\mu_2^2+\nu_2^2}\sin
t_2,\ldots,-\sqrt{\mu_N^2+\nu_N^2}\sin t_N,\nonumber\\&&
\sqrt{\mu_1^2+\nu_1^2}\cos t_1,\sqrt{\mu_2^2+\nu_2^2}\cos
t_2,\ldots,\sqrt{\mu_N^2+\nu_N^2}\cos t_N\Big)\nonumber\\&&
-\sum_{k=1}^N\ln\left(\mu^2+\nu^2\right)=N\ln(\pi e).
\label{T56}\end{eqnarray}

\section{Entropic Inequality for Solitons}\label{sect5}

Entropy of the soliton solution to nonlinear equations was
discussed in \cite{EurPJ-Renato}. In particular, the soliton
solution to Gross--Pitaevskii equation \cite{G-P} was considered
in the tomographic-probability representation to study
Bose--Einstein condensate (BEC) (see also \cite{A1,A2}).

BEC soliton under consideration is given as the function
\begin{equation}\label{Ss1}
\psi(x)=\frac{1}{\sqrt{2l_z}}\,\mbox{sech}\left(\frac{x}{l_z}\right),
\end{equation}
where the parameter $l_z$ describes the soliton width. Symplectic
tomogram of BEC soliton reads
\begin{equation}\label{Ss2}
w_{\rm
S}(X,\mu,\nu)=\frac{1}{2\pi|\nu|}\left|\int\frac{1}{\sqrt{2l_z}}\,
\mbox{sech}\left(\frac{y}{l_z}\right)\exp\left(\frac{i\mu}{2\nu}\,y^2
-\frac{iX}{\nu}\,y\right)\,dy\right|^2,
\end{equation}
where  $\mu=r\cos t$ and $\nu=r\sin t$.

Since $\int|\psi(x)|^2dx=1$, tomogram (\ref{Ss2}) is nonnegative
normalized probability distribution of random position $X$.

The tomographic entropy of BEC soliton equals to
\begin{eqnarray}\label{Ss3}
S(r,t)&=&-\int
\frac{1}{2\pi|\nu|}\left|\int\frac{1}{\sqrt{2l_z}}\,
\mbox{sech}\left(\frac{y}{l_z}\right)\exp\left(\frac{i\mu}{2\nu}\,y^2
-\frac{iX}{\nu}\,y\right)\,dy\right|^2\nonumber\\ &&\times
\ln\left\{ \frac{1}{2\pi|\nu|}\left|\int\frac{1}{\sqrt{2l_z}}\,
\mbox{sech}\left(\frac{y}{l_z}\right)\exp\left(\frac{i\mu}{2\nu}\,y^2
-\frac{iX}{\nu}\,y\right)\,dy\right|^2\right\}\,dX.\nonumber\\ &&
\end{eqnarray}
We introduce the function
\begin{equation}\label{Ss4}
F(r,t)=S(r,t)+S(r,t+\pi/2)-\ln r^2-\ln(\pi e).
\end{equation}
According to the entropic uncertainty relation (\ref{T52}) this
function (we call it entropic uncertainty function) must be
nonnegative. Equation (\ref{T51}) and the additivity property
(\ref{T34}) mean that the entropic uncertainty function (\ref{Ss4})
does not depend on parameter $r$.

Plots of function (\ref{Ss4}) for the Gaussian state and for the
soliton are presented below.

\begin{figure}
\includegraphics[width=0.90\linewidth]{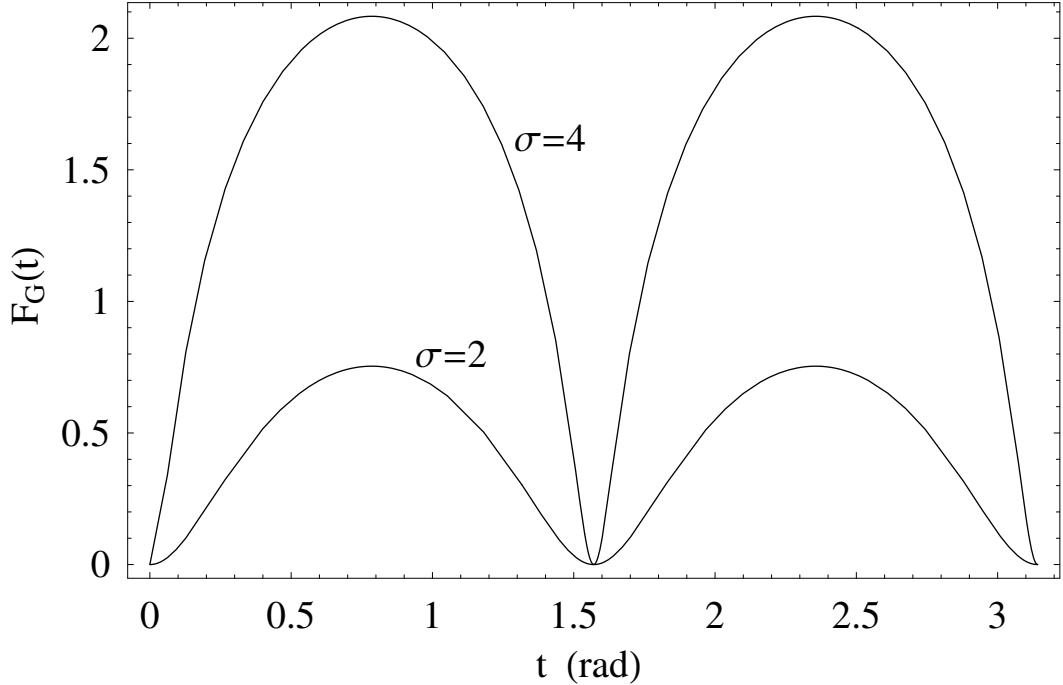}
\caption{\label{figure1}{\it Plot of function $F_{\rm G}(t)$ for
two values of the Gaussian waist $\sigma=2$ and $\sigma=4$.}}
\end{figure}

(i) The normalized initial Gaussian profile is given by
\begin{equation}\label{Ss5}
{\cal F}_{\rm
G}(y)=\frac{\exp(-y^2/2\sigma^2)}{\pi^{1/4}\sigma^{1/2}},
\end{equation}
where $\sigma$ is the waist of Gaussian profile. The corresponding
tomogram calculated with the help of Eq. (\ref{T4}) with $\mu=r\cos
t$ and $\nu=r\sin t$ is given by
\begin{equation}\label{Ss6}
w_{\rm G}(r,t)=\frac{\sigma}{r\sqrt{\pi(\sin^2t+\sigma^4\cos^2t)}}
\exp\left[-\frac{\sigma^2X^2}{r^2(\sin^2t+\sigma^4\cos^2t)}\right].
\end{equation}
The symplectic Gaussian entropy is given as follows:
\begin{equation}\label{Ss7}
S_{\rm G}(r,t)=\frac12-
\ln\left[-\frac{\sigma}{r\sqrt{\pi(\sin^2t+\sigma^4\cos^2t)}}\right].
\end{equation}
The corresponding entropic uncertainty function $F_{\rm G}(t)$ in
this case can be calculated explicitly
\begin{equation}\label{Ss8}
F_{\rm
G}(t)=\ln\left[\sqrt{1+\left(\frac{1-\sigma^4}{2\sigma^2}\right)^2\sin^2
2t}\right].
\end{equation}
Note that the positive definite function $F_{\rm G}(t)$ does not
depend on the radial variable $r$, i.e., it is the same for both
the symplectic and optical entropies and it reduces to zero for
$\sigma=1$, whereas for $\sigma\neq 1$ it is periodic with period
$\pi/2$ as can be seen from FIG.\ref{figure1}.

\begin{figure}
\includegraphics[width=0.90\linewidth]{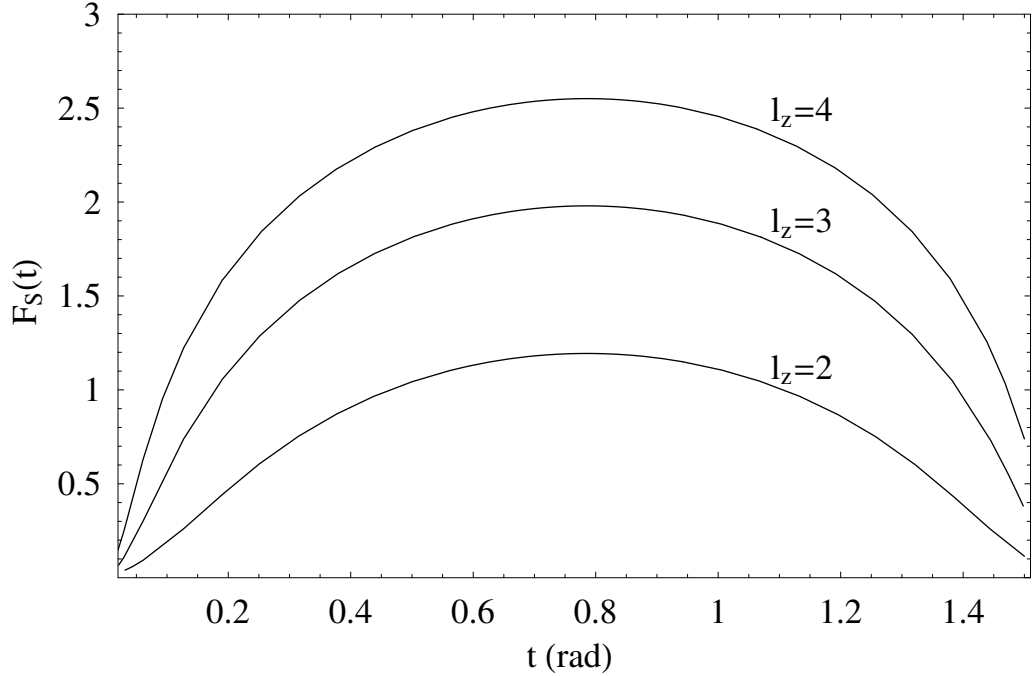}
\caption{\label{figure2}{\it Plot of entropic uncertainty function
$F_{\rm S}(r,t)$ for three values of the soliton-width parameter
$l_z=2$, $l_z=3$, and
$l_z=4$.}}
\end{figure}

(ii) The initial profile of soliton is given by (\ref{Ss1}). Using
the tomographic entropy (\ref{Ss3}), one can calculate the
entropic uncertainty function (\ref{Ss4}) for the soliton
numerically.  The numerical result was obtained by speeding up the
calculation procedure employing fast Fourier transform (FFT)
method. Indeed, it was shown (see \cite{DeNicola}) that tomogram
can be expressed as convolution of the initial profile with a
chirp function (CF). Here the convolution was computed via FFT,
namely, the inverse Fourier transform of the product of FFT of the
initial profile and FFT of CF. Plots in FIG.\ref{figure2}
demonstrate the behaviour of entropic uncertainty function $F_{\rm
S}(t)$ for different values of the soliton-width parameter
$l_z$. One can see that the upper bound of this function
depends on the soliton width.

Before concluding this section, we consider other examples of quantum states with generic Gaussian
Wigner function and the corresponding tomogram
\begin{equation}\label{In1}
w(X,\mu,\nu)=\frac{1}{\sqrt{2\pi\sigma_{XX}(\mu,\nu)}}\exp\left(
-\frac{X^2}{2\sigma_{XX}(\mu,\nu)}\right),
\end{equation}
where \begin{equation}\label{In2}
\sigma_{XX}(\mu,\nu)=\mu^2\sigma_{qq}+\nu^2\sigma_{pp}+2\mu\nu\sigma_{qp}.
\end{equation}
The parameters $\sigma_{qq}$, $\sigma_{pp}$, and $\sigma_{qp}$
satisfy the uncertainty relation
\begin{equation}\label{In3}
\sigma_{qq}\sigma_{pp}-\sigma^2_{qp}\geq 1/4. \end{equation} The
state under consideration for
\begin{equation}\label{In4}
\sigma_{qq}=\sigma_{pp}=\frac{1}{2}\mbox{coth}\,\frac{\beta}{2}\,,
\quad\sigma_{qp}=0
\end{equation}
is the oscillator quantum thermal state with temperature
$T=\beta^{-1}$.

For the state (\ref{In1}), the entropic uncertainty function reads
\begin{eqnarray}\label{In5}
S(t)&=&\ln
2+\frac{1}{2}\ln\left[\sigma_{qq}\cos^2t+\sigma_{pp}\sin^2t+
2\sigma_{qp}\sin t\cos t\right]\nonumber\\ &&\quad
+\frac{1}{2}\ln\left[\sigma_{qq}\sin^2t+\sigma_{pp}\cos^2t-
2\sigma_{qp}\sin t\cos t\right]\,.
\end{eqnarray}

For squeezed thermal state we have
\begin{equation}\label{In6}
\sigma_{qq}=\frac{\lambda}{2}\mbox{coth}\,\frac{1}{2\beta}\,,
\quad
\sigma_{pp}=\frac{1}{2\lambda}\mbox{coth}\,\frac{1}{2\beta}\,,
\quad\sigma_{qp}=0,
\end{equation}
where $\lambda$ is squeezing parameter.

\section{Conclusions}\label{sect6}

To conclude, we point out the main results of this paper.

Inequalities (\ref{T46}) and (\ref{T52}) being the generalizations
of known entropic inequalities for probability distributions of
conjugate position and momentum are obtained for entropies
associated with symplectic tomograms.

The new uncertainty relations obtained characterize the behavior of
quantum state in quantum mechanics as well as the behavior of
analytic signal in signal analysis. The entropic uncertainty
relation for tomographic entropy are obtained also for multimode
quantum state. The uncertainty relation is given by formula
(\ref{T53}). The entropy under study as any Shannon entropy provides
the informational characteristics of the signal.

The nonnegative entropic uncertainty function introduced can be used
to characterize the Shannon information content of a signal, e.g.,
of optical signal.

The uncertainty relation for tomographic entropies is a new
additional property of nonlinear signals including BEC solitons
obeying the Gross--Pitaevskii equation. The physical meaning of
tomographic entropic uncertainty relations will be deepen in a
future work.

\end{document}